\documentclass[a4paper, amsfonts, amssymb, amsmath, reprint, showkeys, superscriptaddress, nofootinbib,oneside, 10pt, aps, prapplied, nobibnotes]{revtex4-2}
\usepackage[english]{babel}
\usepackage[utf8]{inputenc}
\usepackage[normalem]{ulem}
\usepackage{amsmath,bm}
\usepackage{mathtools}
\usepackage{epstopdf}
\usepackage{mathtools} 
\usepackage{extarrows} 
\usepackage{booktabs}
\usepackage{graphicx}
\usepackage{subcaption}
\usepackage[export]{adjustbox}
\usepackage[colorinlistoftodos, color=green!40, prependcaption]{todonotes}
\usepackage{graphicx}
\usepackage[hidelinks]{hyperref}
\usepackage{braket}

\usepackage{enumitem}
\usepackage[linesnumbered,ruled,vlined]{algorithm2e}
\usepackage[noend]{algpseudocode} 
\SetNlSkip{1.0em}

\preprint{APS/123-QED}

\begin{document}
\title{QuantGraph: A Receding-Horizon Quantum Graph Solver}

\author{Pranav Vaidhyanathan}
\thanks{Equal Contribution.\\
\texttt{\{pranav, aristotelis\}@robots.ox.ac.uk}}
\affiliation{Department of Engineering Science, University of Oxford, Oxford OX1 3PJ, United Kingdom}

\author{Aristotelis Papatheodorou}
\thanks{Equal Contribution.\\
\texttt{\{pranav, aristotelis\}@robots.ox.ac.uk}}
\affiliation{Department of Engineering Science, University of Oxford, Oxford OX1 3PJ, United Kingdom}

\author{David R.\ M.\ Arvidsson-Shukur}
\affiliation{Hitachi Cambridge Laboratory, J. J. Thomson Avenue, Cambridge CB3 0HE, United Kingdom}

\author{Mark T.\ Mitchison}
\affiliation{School of Physics, Trinity College Dublin, College Green, Dublin D02 K8N4, Ireland}
\affiliation{Department of Physics, King's College London, Strand, London WC2R 2LS, United Kingdom}
\author{Natalia Ares}
\affiliation{Department of Engineering Science, University of Oxford, Oxford OX1 3PJ, United Kingdom}

\author{Ioannis Havoutis}
\affiliation{Department of Engineering Science, University of Oxford, Oxford OX1 3PJ, United Kingdom}

\begin{abstract}
Dynamic programming is a cornerstone of graph-based optimization. While effective, it scales unfavorably with problem size. In this work, we present QuantGraph, a two-stage quantum-enhanced framework that casts local and global graph-optimization problems as quantum-searches over discrete trajectory spaces. The solver is designed to operate efficiently by first finding a sequence of locally optimal transitions in the graph (local stage), without considering full trajectories. The accumulated cost of the resulting candidate acts as a threshold that prunes the search space (up to $60\%$ reduction for certain examples). The subsequent global stage, based on this threshold, refines the solution. To achieve scalability and robustness, we draw on principles from control theory and embed QuantGraph’s global stage within a receding-horizon model-predictive-control scheme. This classical layer stabilizes and guides the quantum search, improving precision and reducing computational burden. In practice, the resulting closed-loop system exhibits robust behavior and lower query complexity. Notably, for a fixed control horizon and query budget, QuantGraph supports approximately twice as many control discretization bits while still benefiting from Grover-search’s inherent quadratic speedup compared to classical methods.

\end{abstract}
\keywords{quantum-enhanced dynamic programming, two-stage Grover-adaptive-search, quantum-enhanced trajectory optimization, quantum-enhanced model-predictive-control}

\maketitle

\section{Introduction}
\label{sec:intro}

Many decision-making problems in science and engineering can be framed as finding the minimum-cost path through graph-like structures~\cite{blum1997fast, bertsekas2012dynamic, powell_dp}. This simple abstraction underlies applications ranging from autonomous driving and robotics to energy dispatch and logistics. In all such problems, a system must move from one place to another while minimizing effort, risk, or resource use along the way. Classical methods that leverage this structure have long been the workhorse for such problems. Yet, they are challenged by modern applications  that commonly live in big-data regimes and require long planning horizons. 

Robotics provides a particularly compelling illustration of these challenges. Robots have become one of the most transformative technologies of the modern era, supporting society in areas ranging from healthcare and manufacturing to disaster response~\cite{trevelyan2016robotics}, agriculture~\cite{sparrow2021robots}, and home assistance~\cite{yamazaki2012home}. By operating in places that are hazardous, remote, or physically demanding, they extend human reach and enable new forms of service and mobility. Their growing presence in everyday life relies fundamentally on the ability to make rapid and trustworthy decisions in complex, uncertain environments. As these systems take on broader responsibilities, the number of possibilities they must evaluate in real time expands sharply~\cite{papatheodorou2024momentumawaretrajectoryoptimisationusing,vaidhyanathan2025metasymsymplecticmetalearningframework}. This places increasing pressure on planning and decision-making algorithms, underscoring the need for computational tools capable of exploring vast decision spaces both efficiently and reliably.

Quantum computing offers a promising avenue to address these computational bottlenecks~\cite{preskill2018quantum, cerezo2021variational}. Owing to the principles of superposition and entanglement, quantum processors can evaluate many possible transition sequences simultaneously. Algorithms like Grover’s search~\cite{grover1996fastquantummechanicalalgorithm} increase the chance of finding trajectories that meet a chosen metric or constraint, offering an advantage over classical search (see Section~\ref{sec:background}). However, direct implementations that operate on full trajectories require circuits that are too large for current quantum hardware~\cite{shukla2019trajectory}.

Quantum methods have been explored for several robotics tasks.
Chella et al.~\cite{chella2022quantum} use Grover search to select
navigation rules for a robot moving on a grid.
Lathrop et al.~\cite{lathrop2023quantum} apply amplitude amplification
to motion planning, searching for complete paths that avoid obstacles
or for reachable states that extend a rapidly exploring random tree.
Schuetz et al.~\cite{schuetz2022optimization} combine classical
heuristics with quantum annealing for industrial robot trajectory
planning. Their heuristics decode numerical keys into choices that
satisfy the planning constraints.
Other studies focus on robot kinematics.
Salloum et al.~\cite{salloum2025quantum} formulate planar inverse
kinematics as a quadratic optimization problem over binary variables
and solve it on quantum annealing hardware. Their one-hot encoding
uses a separate binary variable for each allowed joint angle.
Nigatu et al.~\cite{nigatu2025quantum} learn a quantum model of
forward kinematics and use Grover search to optimize robot
configurations and design parameters.

These studies show how quantum methods can support navigation, motion
planning, and robot kinematics. In the kinematics examples, quantum
optimization selects a configuration or robot design for a prescribed
task. Controlling a dynamical system also requires choosing a sequence
of inputs: each action changes the state from which later decisions
are made. Searching over complete action
sequences becomes increasingly costly as the task grows longer, while
new measurements may require the plan to be revised during execution.
This motivates methods that account for future motion and incorporate
feedback while limiting the number of qubits needed to represent the
control sequence.

In this work, we introduce QuantGraph, a two-stage quantum-enhanced solver for minimum-cost path problems. QuantGraph is designed to reduce the number of queries needed to search discrete trajectory spaces. The method builds on principles that guide effective decision making in control and planning. It uses receding-horizon feedback and warm-start priors~\cite{mattingley2011receding} to steer the search toward useful regions of the solution space. QuantGraph adapts these ideas to the quantum setting by allocating computation where it matters most. This design aligns with insights from model-predictive-control (see Section~\ref{sec:robotics_background}), reinforcement learning~\cite{SuttonRL} and evolutionary algorithms~\cite{EC}, which show that many real tasks do not require exploring every possible decision to obtain strong solutions.

Building on these principles, the framework operates in two phases. At first, it identifies a small set of promising next actions and evaluates the resulting candidate using the global objective. This produces a baseline that serves as a warm-start prior. Then, the quantum oracle marks trajectories that improve on this baseline. It does so by considering the full-planning horizon of the problem. QuantGraph focuses computation on the most relevant parts of the search space instead of examining all possible paths.

To ensure scalability beyond current quantum-search methods, the second stage of QuantGraph follows a receding-horizon strategy~\cite{kouvaritakis2016model}. Instead of solving the entire problem at once, it optimizes a short control window, applies the first action, and then shifts the window forward as new information becomes available. This approach is standard in control theory and allows the solver to correct errors and adapt to changing conditions at each replanning step. As the window advances, each iteration begins from the updated state and uses a newly refined threshold. This allows QuantGraph to focus its resources on the most relevant region of the search space at that moment.

To demonstrate the capabilities of our proposed framework, we focus on the challenging problem of trajectory optimization for discrete-time dynamical systems. In such problems, a system evolves along a trajectory represented as a path through a temporal graph of states (nodes) and actions (edges). By recasting the underlying problem into a quadratic-unconstrained-binary-optimization formulation~\cite{Qubo}, we are able to utilize QuantGraph's scalability and inherent quadratic speedup with profound implications across many scientific fields. Section~\ref{sec:results} presents our findings in detail.

\subsection{Background}
\label{sec:background}

To frame the operation of QuantGraph, we begin with a concise overview of the core ideas and notation used by the robotics and quantum computing communities. The aim is to build an intuitive understanding of the underlying principles without resorting to formal derivations.

\subsubsection{Robot dynamics and trajectory optimization}\label{sec:robotics_background}

A core problem in robotics is to drive a robot, or more generally, a dynamical system, toward a desired goal. Among model-based approaches, trajectory optimization provides a systematic framework to achieve this. Such methods rest on two fundamental pillars: the \textbf{dynamics model}, which predicts how the system evolves under control inputs, and the \textbf{optimization solver}, which determines the sequence of control inputs that minimize a surrogate objective describing the task or goal given the system's dynamics.

\paragraph*{\textbf{Dynamics modeling:}}

A robot’s motion evolves according to physical laws that relate its state ($\mathbf x_k$) (e.g., positions, velocities, orientations) and \textbf{control input} ($\mathbf u_k$) (e.g., motor torques, wheel forces) across time. In discrete time, the state-evolution for a time step ($k$) is described by the state-space model~\cite{siciliano2008springer},
\begin{equation}
    \mathbf x_{k+1} = f(\mathbf x_k, \mathbf u_k),
    \label{nonlinear_dynamics}
\end{equation}
where $f(\cdot,\cdot)$ can be nonlinear. For linear time invariant dynamics, or after linearization about a fixed equilibrium in deviation coordinates, we obtain the \textbf{linear time-invariant} form~\cite{willems1986time, oppenheim1997signals},
\begin{equation}
    \mathbf x_{k+1} = \mathbf A \mathbf x_k + \mathbf B  \mathbf u_k,
    \label{LTI}
\end{equation}
which captures many robotic systems such as manipulators or wheeled platforms near equilibrium \cite{sastry2011adaptive}. These dynamics define how the system moves from a state to the next, and thus determine the transitions that form a trajectory. The collection of all possible trajectories can be visualized as a graph. Each node represents a state ($\mathbf x_k$), and each edge represents the transition to the next state ($\mathbf x_{k+1}$). By encoding the problem with this graph structure, we can represent every admissible path the system can follow over time.

\begin{figure*}[ht!]
  \centering
  \begin{subfigure}[t]{0.52\textwidth}
    \centering
    \includegraphics[valign=t,width=\linewidth]{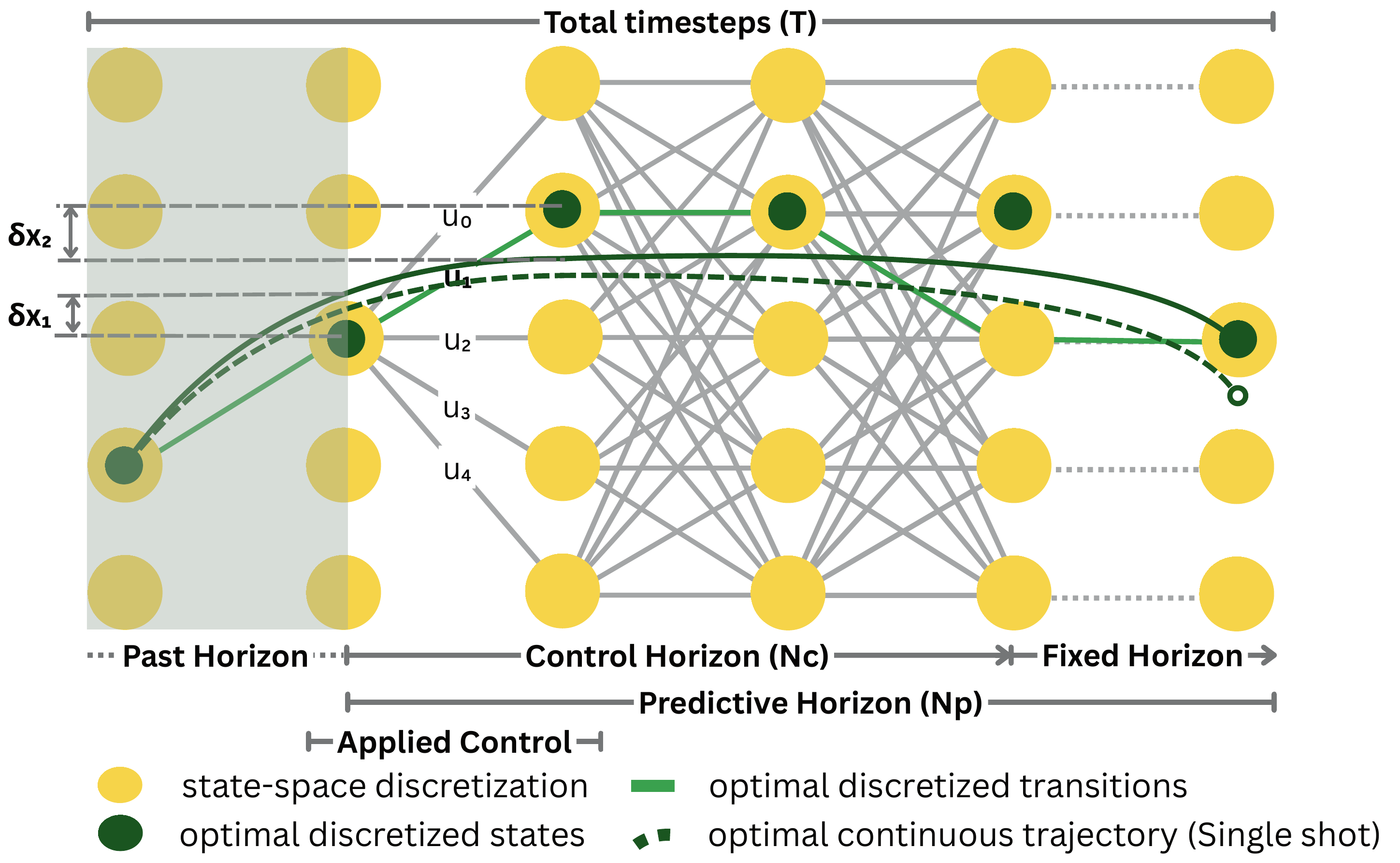}
    \caption{Iteration 1 of the model predictive controller}
    \label{fig:MPC_iter1}
  \end{subfigure}
  \hfill
  \begin{subfigure}[t]{0.47\textwidth}
    \centering
    \includegraphics[valign=t,width=\linewidth]{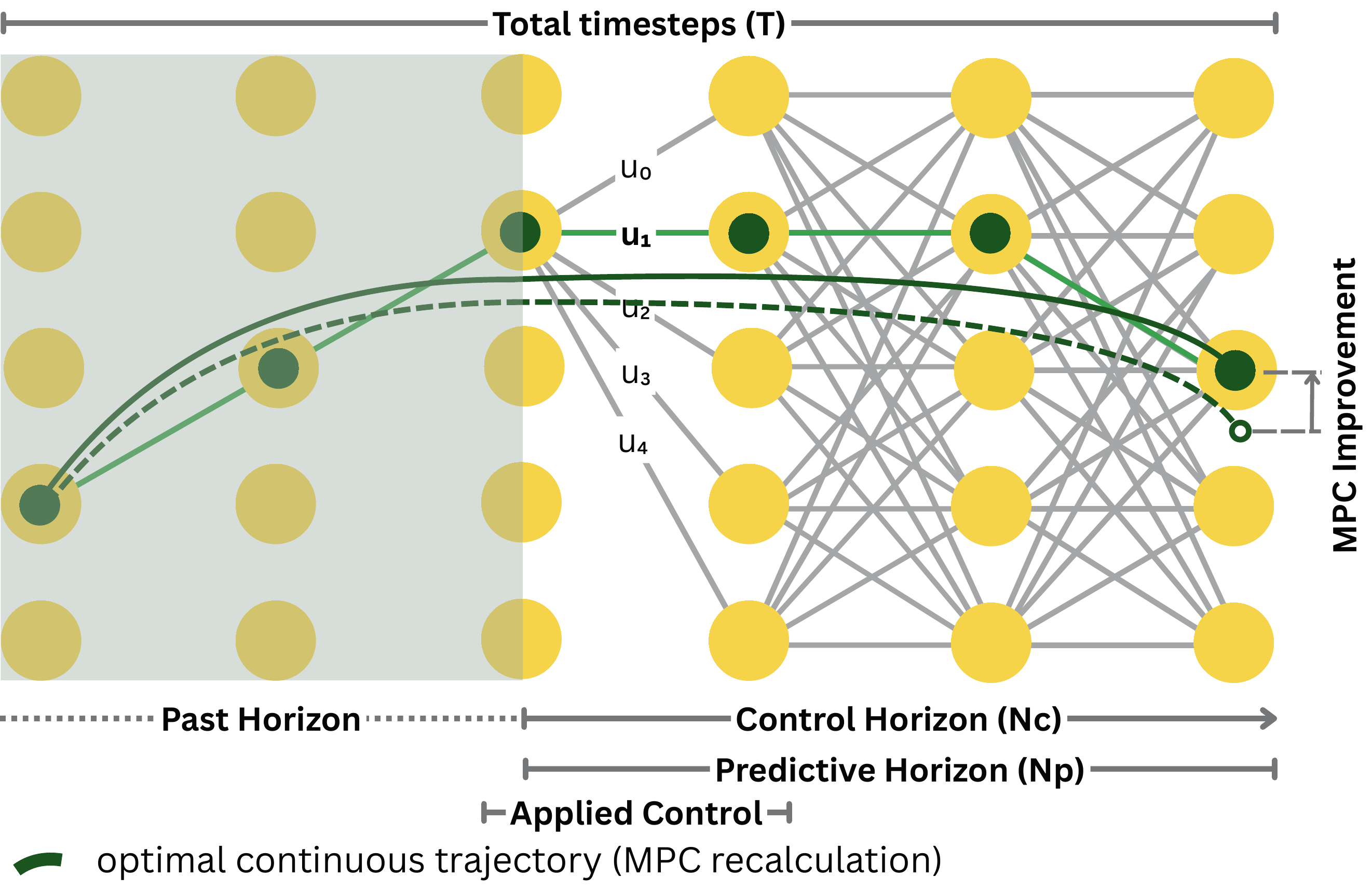}
    \vspace{3mm}
    \caption{Iteration 2 of the model predictive controller}
    \label{fig:MPC_iter2}
  \end{subfigure}
  
  \caption{Subsequent model-predictive-control (MPC) iterations: (a) In early model-predictive-control steps, discretized states deviate from the continuous optimum, with compounding errors $\delta x_1 < \delta x_2$ along the horizon. To limit computation over the full horizon $T$, the model-predictive-control framework optimizes only the first $N_c \ll T$ control actions, while beyond $N_c$ the control-inputs are held constant (the \emph{Fixed Horizon} block). Holding the remaining inputs fixed reduces computation but limits the
choice of future trajectories. Only the first control input is applied;
the next optimization starts from the updated state and can adjust the
remaining plan to compensate for earlier deviations. (b) In subsequent iterations, the model-predictive-control framework recomputes controls from the updated state, compensating for prior deviations and refining the trajectory.}
  \label{fig:MPC}
\end{figure*}

\paragraph*{\textbf{Trajectory optimization:}}

The goal of trajectory optimization is to find the optimal control sequence $\{\mathbf u^*_k\}_{k=0}^{T-1}$ which, given the system dynamics in Eq.~\eqref{nonlinear_dynamics}, steers the system over a finite horizon of length $T$~\cite{rao2014trajectory,betts1998survey}. This horizon can be represented as a $T$-layered tree, where each layer $k=0,\dots,T$ collects all states ($\mathbf x_k$) that are reachable at time step $k$, and each edge between layers corresponds to applying an admissible control input ($\mathbf u_k$) and propagating the dynamics to the next state ($\mathbf x_{k+1}$) (see Fig.~\ref{fig:MPC}). Any root-to-leaf path in this tree is thus a dynamically feasible trajectory. Among all such trajectories, we seek the one that minimizes an accumulated performance index,
\begin{equation}
    J = \sum_{k=0}^{T-1} \ell_k(\mathbf x_k, \mathbf u_k) + \phi(\mathbf x_T),
    \label{eq:accumulated_cost}
\end{equation}
where $\ell_k(\mathbf x_k, \mathbf u_k)$ is the \textit{stage-wise cost} and $\phi(\mathbf x_T)$ is the terminal penalty. Common quadratic cost functions~\cite{mastalli20crocoddyl} are,
\begin{equation}
\begin{aligned}
    \ell_k(\mathbf x_k, \mathbf u_k) &= (\mathbf x_k - \mathbf x_k^{\mathrm{ref}})^{\top} \mathbf Q (\mathbf x_k - \mathbf x_k^{\mathrm{ref}}) \\
    &\quad + (\mathbf u_k - \mathbf u_k^{\mathrm{ref}})^{\top} \mathbf R (\mathbf u_k - \mathbf u_k^{\mathrm{ref}}),\\
    \phi(\mathbf x_T) &= (\mathbf x_T - \mathbf x_T^{\mathrm{ref}})^{\top} \mathbf P (\mathbf x_T - \mathbf x_T^{\mathrm{ref}}),
    \label{eq:cost}
\end{aligned}
\end{equation}
where $\mathbf Q, \mathbf R, \mathbf P \succeq 0$ are weighting matrices that balance tracking accuracy against control effort. Moreover, $\mathbf x^{\mathrm{ref}}, \mathbf u^{\mathrm{ref}}$ denote a reference trajectory that regularizes the problem and encourages smooth convergence of the solver to a desirable motion. Intuitively, the quadratic terms penalize deviations from the reference state and input at each time step, while the terminal cost pushes the system to reach the intended terminal state. Minimizing $J$ therefore selects, among all feasible paths in the $T$-layered tree, the dynamically consistent trajectory that remains close to the desired motion while using control inputs efficiently. In the numerical examples, the stage costs penalize deviations from the target state and reference input, while the terminal cost further penalizes deviation from the goal.

\paragraph*{\textbf{Dynamic programming:}}

The trajectories or transitions in the aforementioned graphs are evaluated recursively by algorithms such as classical dynamic programming~\cite{DP_2010}, which relies on Bellman's principle of optimality. Intuitively, the principle states that any suffix of an optimal trajectory must itself be optimal for the corresponding intermediate state. Formally, the optimal cost-to-go $J_k^*(\mathbf x_k)$ satisfies
\begin{equation}
\begin{aligned}
    &J_k^*(\mathbf x_k) = \min_{\mathbf u_k}\big[\ell_k(\mathbf x_k,\mathbf u_k) + J_{k+1}^*(f(\mathbf x_k,\mathbf u_k))\big],\\
    &J_T^*(\mathbf x_T)=\phi(\mathbf x_T),
\end{aligned}
\end{equation}
which expresses the decomposition of the global-optimization problem into a sequence of stage-wise decisions. The recursion states that each optimal action minimizes its immediate cost plus the best future cost resulting from that action. This allows efficient backward traversal over the admissible state transitions and guarantees a globally optimal trajectory. However, global optimality comes at a cost. Even with this recursion, dynamic programming suffers from the curse of dimensionality. For highly combinatorial tasks, its complexity can grow exponentially with problem size, making real-time control impractical.

\paragraph*{\textbf{Model predictive control:}}

This challenge can be mitigated by solving a \textit{finite-horizon} optimal control problem at each time step. We distinguish the \emph{predictive horizon} $N_p$ from a shorter \emph{control horizon} ($N_c$) and the total number of time steps of the task ($T$), with $N_c \le N_p \leq T$. Over $N_p$ steps we predict states and accumulate cost, but we only optimize the first $N_c$ control actions. 
Beyond $t + N_c - 1$, the inputs are constrained by a simple \emph{terminal (tail) policy}, for example, applying the last optimized input $\mathbf u_{t + N_c - 1}$.  This \emph{fixed} tail reduces decision variables while the receding-horizon recalculations allow corrective replanning from the updated state (see Figure~\ref{fig:MPC}).

Starting from the current state ($\mathbf x_t$), the controller minimizes over the stage-wise control inputs ($\mathbf u_k$) the following problem:
\begin{equation}
\begin{aligned}
    \min_{\{\mathbf u_k\}_{k=t}^{t+N_c-1}} &\sum_{k=t}^{T-1} \ell_k(\mathbf x_k, \mathbf u_k) + \phi(\mathbf x_{T}),\\
    &\quad \text{s.t.} \quad \mathbf x_{k+1}=f(\mathbf x_k,\mathbf u_k)\\
    &\mathbf u_k = 
    \begin{cases}
\mathbf u_k, & k \le t + N_c - 1, \\
\mathbf u_{t + N_c - 1}, & k \ge t + N_c,
\end{cases}
    \label{eq:MPC_objective}
\end{aligned}
\end{equation}
over a shorter \textbf{control horizon} ($N_c \ll T$). Only the first optimal control ($\mathbf u_t^\star$) is applied to the system, after which new measurements update the state and the optimization repeats (i.e. \textit{receding-horizon} principle). By continually re-optimizing in a sliding-window fashion with fresh data, model-predictive-control naturally adapts to model errors, disturbances, and changing goals. This, effectively, provides a feedback mechanism that bridges optimal planning and real-time control. Figure~\ref{fig:MPC} illustrates the underlying operation principles~\cite{Kober2014,SuttonRL}.

In this work, we reinterpret trajectory optimization and model-predictive-control through a \textbf{quantum-enhanced lens}, framing the problem as a graph search over discrete trajectories. This allows us to leverage quantum-search algorithms to achieve scalable, efficient, and adaptive control for complex robotic systems.

\subsubsection{Quantum search and optimization}\label{sec:quantum_background}

The limits of classical dynamic programming and large-scale graph searches motivate the use of quantum-search algorithms, which can explore large discrete spaces more efficiently. To clarify this connection, we briefly introduce the notation and principles underlying quantum computation.

A quantum computer encodes information in a \emph{state vector} $|\psi\rangle$ that resides in a complex Hilbert space $\mathcal{H} \cong \mathbb{C}^{N}$, where $N = 2^{M}$ is the size of the computational basis for a register of $M$ qubits. This state is a normalized superposition of basis vectors,
\[
|\psi\rangle = \sum_{x=0}^{N-1} \psi_x |x\rangle, \qquad \sum_{x=0}^{N-1} |\psi_x|^2 = 1,
\]
with each complex amplitude $\psi_x$ representing the contribution of basis state $|x\rangle$~\cite{nielsen2010quantum}. In our setting, each basis state can be viewed as encoding one discrete trajectory through a graph.

Quantum computation proceeds through the application of unitary operators $U$, which play a role analogous to reversible transition maps in classical systems. Because $U^\dagger U = I$, these operators preserve norms and ensure deterministic state evolution:
\[
|\psi_{\mathrm{final}}\rangle = U\,|\psi_{\mathrm{initial}}\rangle.
\]

Measurement is where quantum mechanics departs most clearly from classical computation. A classical register returns a definite value. On the contrary, a quantum state must be sampled probabilistically. Upon measurement, the outcome is a bitstring $x$ drawn with probability
\[
P(x) = |\langle x | \psi \rangle|^2,
\]
according to the Born rule~\cite{nielsen2010quantum}. Thus, while the computation evolves deterministically, the readout is inherently random, reflecting the amplitudes accumulated along different encoded trajectories.

\paragraph*{\textbf{Grover's algorithm:}}

Grover's quantum search algorithm is a cornerstone of quantum optimization, offering a provable quadratic speedup for unstructured-search tasks~\cite{grover1996fastquantummechanicalalgorithm}. Whereas a classical search through an unsorted database of size N requires $O(N)$ queries, Grover's algorithm can identify a marked element in just $O(\sqrt{N})$ queries. The algorithm's power stems from three core quantum-mechanical steps:

\begin{enumerate}
    \item \textbf{State preparation:} The process begins by preparing an $n$-qubit register, where $N=2^n$, in a uniform superposition of all possible computational basis states. This is typically achieved by applying a Hadamard gate to each qubit in the $\ket{0}^{\otimes n}$ state, yielding the state $\ket{s} = \frac{1}{\sqrt{N}}\sum_{x=0}^{N-1}\ket{x}$~\cite{nielsen2010quantum}. This initial state represents all candidate solutions simultaneously with equal amplitude. In the context of trajectory optimization, this superposition encodes the entire search space of possible discretized control sequences (trajectories) the dynamical system can execute over the planning horizon.
    
    \item \textbf{The quantum oracle ($\mathcal{O}$):} The heart of the algorithm is an oracle, which is a unitary operator designed to recognize solutions. For a search problem defined by a function $m(x)$ that returns 1 for a solution state $\ket{x_w}$ and 0 otherwise, the oracle applies a conditional phase shift, such that $\ket{x} \mapsto (-1)^{m(x)}\ket{x}$. This operation effectively \emph{marks} the solution state by inverting its phase (e.g., $\ket{x_w} \mapsto -\ket{x_w}$) while leaving all other states unchanged. Crucially, the oracle does this without collapsing the superposition~\cite{kashefi2002comparison}.
    \item \textbf{Amplitude amplification ($\mathcal{D}$):} Following the oracle's application, a diffusion operator, $\mathcal{D} = 2\ket{s}\bra{s} - I$ (where $I$ is the identity), is applied. This operator performs an inversion about the mean of the amplitudes of all states. Geometrically, this operation amplifies the amplitude of the marked state (which has a negative phase) while shrinking the amplitudes of all other states.
\end{enumerate}

The sequential application of the oracle and the diffusion operator constitutes a single \emph{Grover iteration}. For a single marked state, the required number of iterations grows approximately as $\frac{\pi}{4}\sqrt{N}$, and the probability of measuring that state approaches one as $N$ increases.

\paragraph*{\textbf{Grover-adaptive-search:}}

While the canonical algorithm is designed for finding a state marked by an oracle, it can be adapted for optimization through methods such as Grover-adaptive-search~\cite{Gilliam_2021}. Grover-adaptive-search reframes the optimization problem as a series of quantum searches. It iteratively searches for any solution with a cost below a dynamically updated threshold, $\tau$. Upon finding a superior solution, the threshold is lowered to the new best-known cost, progressively converging toward the global minimum. This transforms Grover's algorithm from a simple search primitive into a powerful minimum-finding solver.

\paragraph*{\textbf{Implications and challenges:}}

The implications of this quadratic speedup are profound for discretized continuum problems, such as the trajectory optimization tasks central to robotics. For a fixed-query budget, a quantum search can interrogate a quadratically larger solution space than its classical counterpart. As shown in Section \ref{sec:analysis}, the quantum advantage allows for a doubling of the bits of precision ($M_{quantum} \approx 2 \cdot M_{classical}$) in control discretization. Effectively, this translates to a quadratic reduction in normalized control grid spacing.

\subsection{Overview of our approach}\label{sec:contributions}
In what follows, we describe QuantGraph's main components and capabilities. Primarily, we introduce a quantum-enhanced solver for generic graph optimization problems, with a particular focus on minimum-cost path search. This reframes trajectory optimization problems that are typically addressed via dynamic programming into a more general graph-search setting with a quadratic-unconstrained-binary-optimization formulation~\cite{chella2022quantum}.

Our framework applies Grover-adaptive-search to this formulation, inheriting its quadratic advantage in search scaling over dynamic programming on the fully expanded trajectory tree at fixed control and prediction horizons. The advantage here is significant, since instead of evaluating the entire decision tree, the quantum search directly promotes low-cost trajectories. As a result, the global task of identifying high-quality paths from an exponentially large set can be carried out without exhaustive enumeration. For a fixed query budget, this allows the solver to explore more paths with finer discretizations than this classical baseline.

QuantGraph is organized as a two-stage, local–global architecture. The local stage considers only a single-step horizon that requires fewer qubits to quickly identify promising immediate actions. The global cost of the resulting candidate serves as a threshold that warm-starts the global Grover-adaptive-search. The global stage then refines the solution using this initial threshold, which can accelerate the discovery of lower cost trajectories.

Finally, we embed QuantGraph within a receding-horizon model-predictive-control framework to manage large, dynamically-evolving state spaces. Instead of constructing and searching the full set of possible trajectories at once, the solver optimizes over a sliding window, substantially reducing memory demands. As shown in Section~\ref{sec:analysis}, this design converts the exponential query complexity of the original Grover-adaptive-search algorithm into a linear dependence on the task horizon for a fixed window size and control discretization. This in turn yields finer control grids within a fixed query budget.

\section{Results}\label{sec:results}
We evaluate QuantGraph on a range of graph-based traversal and path-planning problems. We begin with a static example to illustrate the roles of the local and global stages and their effect on the initial search threshold. We then examine more challenging, robotics-inspired settings involving systems with dynamically evolving trajectories, such as the linear time-invariant double integrator and the highly nonlinear cart pole. Each benchmark highlights different strengths and limitations of the framework. Quantum search is simulated on classical computers using IBM's Qiskit~\cite{qiskit}. No physical quantum processor is used. Appendix~\ref{app:numerical_setup} describes the illustrative implementations and their numerical settings.

\subsection{Static graph}\label{sec:static_graph}
\begin{figure}[htb!]
    \includegraphics[width=\linewidth]{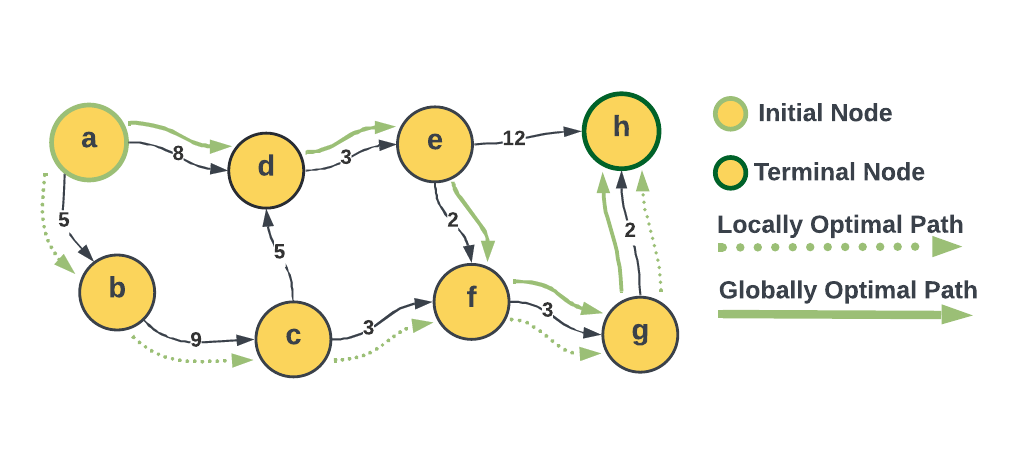}
    \caption{A simple navigation problem as a directed graph with weighted transitions. QuantGraph is called to find the path associated with the lowest cost possible from node \textbf{a} to node \textbf{h}. The local-search step provides a candidate whose cost sets the initial threshold for the global search.}
    \label{fig:static_graph}
\end{figure}

To illustrate how our approach operates, we begin with a small, static navigation problem.
This problem is described by the static directed-acyclic-graph illustrated in Figure~\ref{fig:static_graph}. The graph consists of nodes connected with transitions. Each transition has an associated cost. The task is to traverse the graph and accumulate the lowest cost possible from node~\textbf{a} to node ~\textbf{h}.

In our example, there are five possible paths that the solver could take. Let $\{\mathbf a, \dots, \mathbf h\} \in \mathcal X$ represent a trajectory inside the set of all feasible trajectories $\mathcal X$. The cost associated for each trajectory is shown in Table~\ref{tab:static_graph_traj_cost}. The local candidate has a lower cost than three of the five trajectories and sets the initial threshold for the global search ($60\%$ search space reduction). This initialization may also be useful for larger graphs, such as those found in social media platforms~\cite{facebook_anatomy}.
\begin{table}[ht]
  \centering
  \begin{tabular}{@{} l | r @{}}
    \toprule
    Trajectories & Cost \\
    \hline\hline
    \midrule
    $\{\mathbf a, \mathbf d, \mathbf e, \mathbf h\}$ &  23 \\
    $\{\mathbf a, \mathbf d, \mathbf e, \mathbf f, \mathbf g, \mathbf h\}$ & \textbf{18} \\
    $\{\mathbf a, \mathbf b, \mathbf c, \mathbf f, \mathbf g, \mathbf h\}$ & 22 \\
    $\{\mathbf a, \mathbf b, \mathbf c, \mathbf d, \mathbf e, \mathbf h\}$ & 34 \\
    $\{\mathbf a, \mathbf b, \mathbf c, \mathbf d, \mathbf e, \mathbf f, \mathbf g, \mathbf h\}$ & 29 \\
    \bottomrule
  \end{tabular}
  \caption{List of all possible trajectories from Figure \ref{fig:static_graph} and their associated costs.}
  \label{tab:static_graph_traj_cost}
\end{table}

\subsection{Double integrator}\label{sec:double_int}
We evaluate QuantGraph on a highly combinatorial, dynamically evolving graph associated with trajectory optimization problems. More specifically, the problem concerns a discretized linear time-invariant double-integrator system. For example, the system can be thought of as a box of unit mass ($1$ kg) on a frictionless one-dimensional plane. Its dynamics are described in Eq.~\eqref{eq:double_int}, in which $q$ represents the position of the system and $u$ the driving force~\cite{ren2008consensus}, which is bound constrained by the maximum admissible force $u_{max}$. Each time step of the discretized model is represented by the index $k$.
\begin{figure*}[ht!]
    \includegraphics[width=1.0\textwidth]{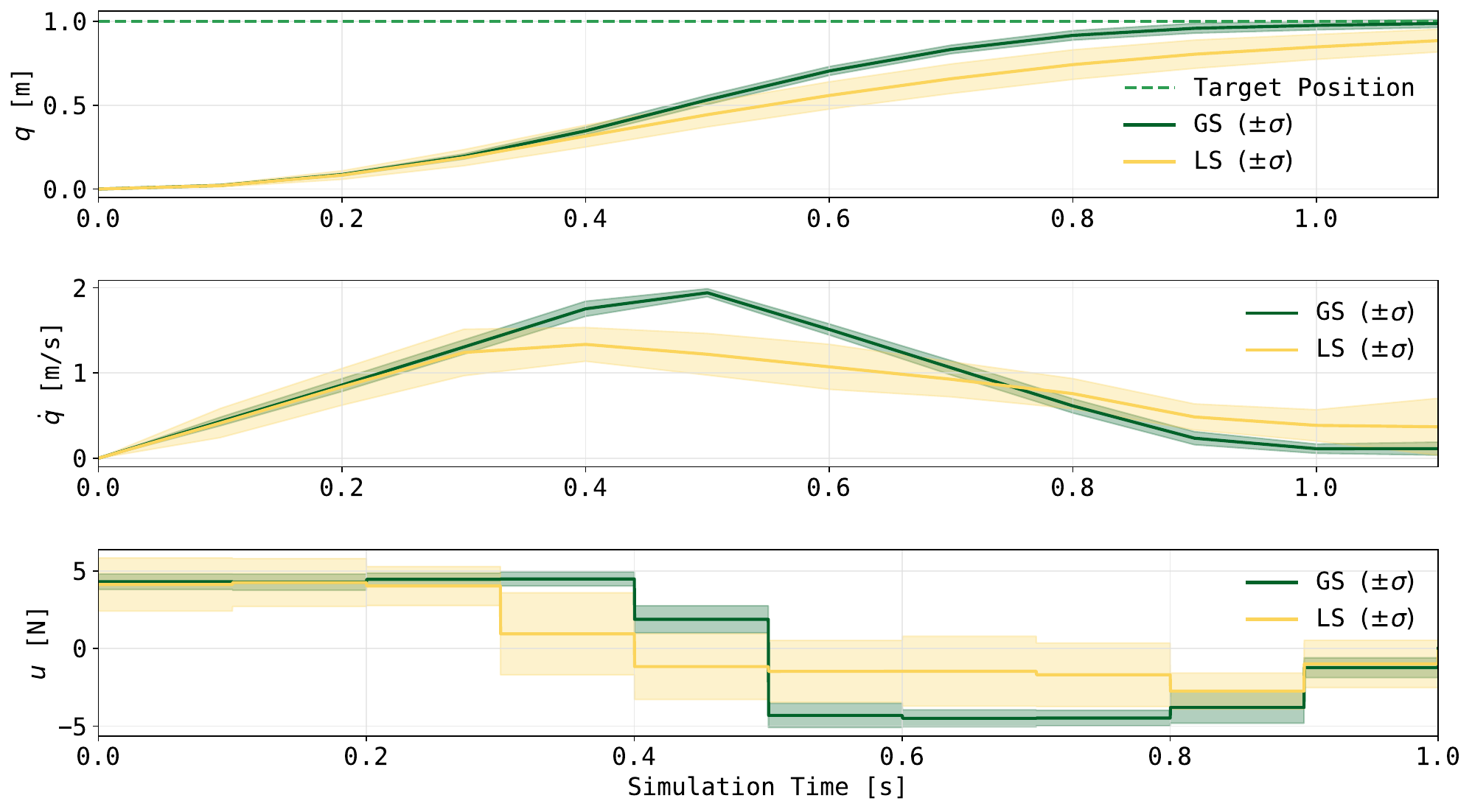}
    \caption{Planned trajectories for the double integrator: The local stage warm starts the global stage of QuantGraph that operates in a receding horizon fashion. The objective is to drive the double integrator to the target position. To demonstrate the consistency of our framework, we plot the results for ten runs. The local stage converges to a sub-optimal trajectory as expected, warm-starting the global stage that converges to a smooth trajectory with low variance across runs. }
    \label{fig:double_integrator}
\end{figure*}
\begin{equation}
\begin{aligned}
    \ddot{q}&=u(t), \quad|u(t)| \leq u_{\text{max}} \Rightarrow \\
    \dot{\mathbf x} &= \overbrace{\begin{bmatrix}  0 & 1 \\0 & 0\end{bmatrix}}^{\mathbf A}\mathbf x + \overbrace{\begin{bmatrix} 0 \\ 1\end{bmatrix}}^{\mathbf B}u(t) \text{, with } \mathbf x=\begin{bmatrix} q \\ \dot{q} \end{bmatrix} \Rightarrow \\
    \mathbf x_{k+1} &= e^{\mathbf Adt}\mathbf x_k + \int_{t_k}^{t_k+dt}e^{\mathbf A\left( t_k+dt-\tau\right)}\mathbf Bu\; d\tau \Rightarrow \\
    \mathbf x_{k+1} &= \begin{bmatrix}  1 & dt \\0 & 1\end{bmatrix}\mathbf x_k + \begin{bmatrix} \frac{1}{2}dt^2 \\ dt\end{bmatrix}u_k.\\
\end{aligned}
\label{eq:double_int}
\end{equation}

In terms of complexity the dynamics are linear, but the set of feasible trajectories forms a large, dynamically evolving tree. QuantGraph’s receding-horizon global stage mitigates this problem by enabling feedback control without searching the full tree. As in classical model-predictive-control, short horizons may yield solutions that are not fully optimal, but the full remaining horizon problem is recovered when $N_c=N_p=T-t$. The trade-off between tractability and full-horizon optimality is inherent to large graph optimization problems, where memory limits naturally motivate a sliding-window approach. The results for the double integrator example are illustrated in Figure~\ref{fig:double_integrator}.

\subsection{Cart pole}\label{sec:cartpole}
We evaluate our framework on a cart-pole trajectory-optimization task. In this problem, an underactuated pole with a mass is attached on top of an actuated cart moving freely on a frictionless one-dimensional plane. The goal is to raise the pole to a standing position. The system is described in Eq.~\eqref{eq:cart_pole}, where $m$ represents the mass ($0.1$ kg) ontop of the pole ($l$) with length $0.5$ m, $M$ the cart's mass ($1$ kg) and $F$ the force driving the cart~\cite{geva2002cartpole}:
\begin{figure*}[ht!]
  \centering
  \begin{subfigure}[t]{0.68\textwidth}
    \centering
    \includegraphics[width=\linewidth]{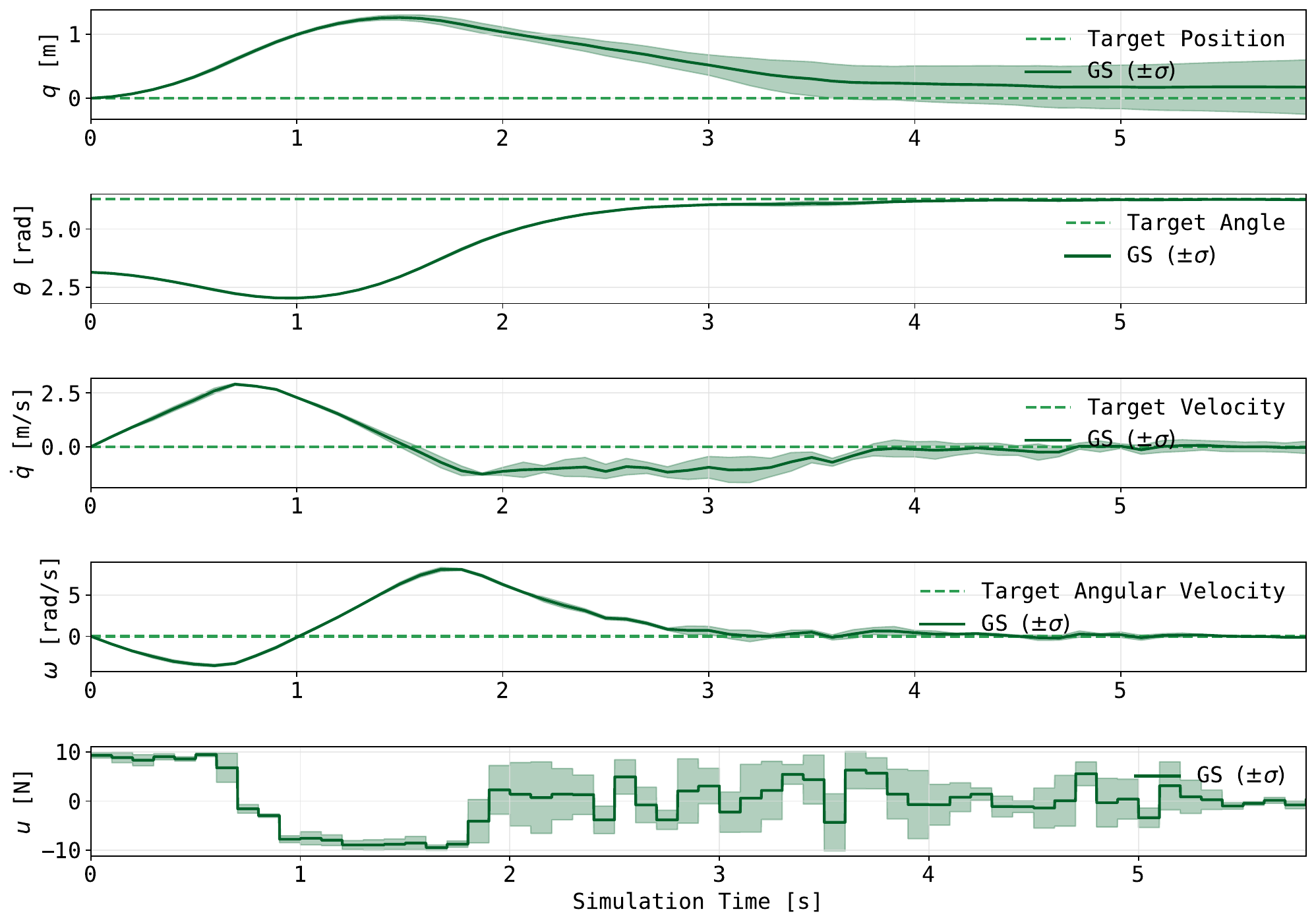}
    \caption{}
    \label{fig:planned_trajectories}
  \end{subfigure}
  \hfill
  \begin{subfigure}[t]{0.31\textwidth}
    \centering
    \includegraphics[width=\linewidth]{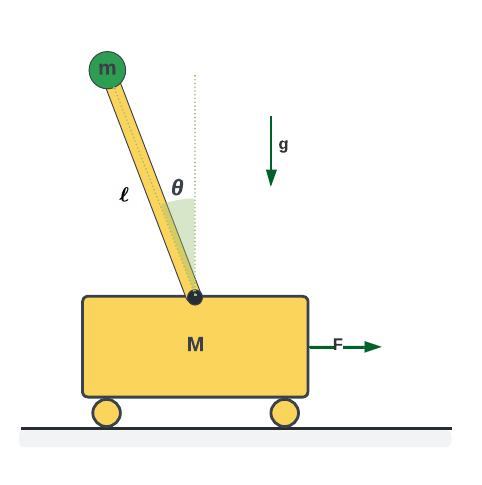}
    \caption{}
    \label{fig:system}
  \end{subfigure}
    \caption{(a) Planned trajectories for the cart pole system. The primary objective is to swing up the underactuated pole mass to the upright position while minimizing displacement of the cart from its initial location (secondary objective). The plot shows the mean and standard deviation over ten independent runs. The negligible variance in the evolution of the pole angle $\left(\theta\right)$ highlights the consistency and reliability of our planning approach. (b) Schematic of the physical cart pole system, annotated with the variables used in Eq.~\eqref{eq:cart_pole}.}
  \label{fig:cartpole}
\end{figure*}

\begin{equation}
\begin{aligned}
    &\ddot{\theta} = \frac{1}{\mu(\theta)} \big( \frac{\cos \theta}{l} F + \frac{(m+M)g}{l} \sin(\theta)\\&\qquad \qquad \qquad- m \cos(\theta) \sin(\theta) \dot{\theta}^2\big),\\
    &\ddot{q} = \frac{1}{\mu(\theta)} \big( F + m \cos(\theta) \sin(\theta) g - m l \sin(\theta) \dot{\theta}^{2} \big),
\end{aligned}
\label{eq:cart_pole}
\end{equation}
and where $\mu(\theta) := M + m \sin(\theta)^2$ designates the effective mass of the cart pole.
Here, $\theta$ is measured counterclockwise from the
upward vertical, while positive $q$ and $F$ point to the right.
The nonlinearity of this system makes it a meaningful benchmark for assessing QuantGraph’s capabilities. The local stage, designed for linear time-invariant systems with quadratic costs, is less effective in handling the cart pole's nonlinearities. Future extensions could relax this limitation by incorporating ancilla qubits~\cite{AncillaQubitsRef}, though at the cost of increased circuit complexity. This in turn will provide efficient warm-start capabilities, accelerating the convergence of the overall algorithm. 

On the other hand, QuantGraph’s global stage is inherently capable of addressing general nonlinear dynamics and non-quadratic cost landscapes. This allows it to successfully solve challenging control tasks such as the cart-pole problem, even in the absence of guidance from the local stage, thereby underscoring the versatility of our approach.

The results for the cart-pole, path-planning problem are presented in Figure~\ref{fig:cartpole}. In this numerical demonstration, the solver's global stage is initially cold-started, with each subsequent iteration warm-started by the shifted previous solution.

More specifically, the global stage operates directly on the cost function and system’s native nonlinearities. By executing this procedure in a receding-horizon manner, our framework replans from the current state to guide the pole towards the upright position. This highlights its effectiveness in handling complex, nonlinear control tasks even without the local stage.

\section{Discussion}
\label{conclusion}
In this work, we introduced QuantGraph, a quantum-enhanced framework for solving graph-structured optimization problems, including trajectory planning in robotics and control. By reformulating dynamic programming as quantum search over discrete-trajectory spaces, we identified a quadratic advantage in oracle queries (see Section~\ref{sec:analysis}). Our two-stage architecture uses Grover-adaptive-search at its core. As shown in Section~\ref{sec:analysis}, our framework delivers the expected quadratic speedup while remaining computationally tractable. Empirical results on the double-integrator and cart-pole systems demonstrate that QuantGraph performs well on both linear and nonlinear dynamics.

QuantGraph can also search graphs that do not arise from robot dynamics.
A candidate path can be encoded through its nodes, edges, or branching
choices. A reversible oracle then computes the total edge cost and checks
that the path connects the required source and target through valid
transitions, together with any additional constraints. This construction
can accommodate directed or undirected, sparse or dense, and static or
dynamically generated graphs, provided that candidate paths have finite
encodings and their costs and constraints can be evaluated reversibly.
The cost of accessing graph data and evaluating these quantities must
also be included when assessing computational resources. This
representation does not establish an advantage over every classical
shortest-path algorithm.
The control problem formulated in Section~\ref{sec:methods} is one
structured example. Appendix~\ref{app:constraints} explains which
constraints are satisfied by the encoding and how others could be
included as cost penalties.

The benefit of the receding-horizon approach depends on the length of
the paths and whether decisions can be revised as the traversal proceeds.
Sparsity alone does not determine the difficulty: even a graph with few
branches at each node can contain exponentially many paths as their
length increases. Optimizing a shorter sequence reduces the number of
qubits needed to encode the decisions and allows the path to be updated
during execution. For a small static graph that can be searched in full,
a single optimization or a classical shortest-path algorithm may be
preferable. Restricting the future control sequence reduces the search
space but need not preserve the optimum of the unrestricted problem.
The linear dependence on task duration in Section~\ref{sec:analysis}
concerns oracle-query counts for fixed $N_c$ and $M$; the arithmetic within
each query can still depend on the prediction horizon.

A warm start can be used whether the global stage optimizes the full
trajectory or a shorter control window. A feasible trajectory from the
local stage or a classical heuristic can provide an initial candidate.
It must first be represented on the current control grid and satisfy
any restriction on the remaining inputs, including the fixed tail.
Its cost is then evaluated using the global objective
to obtain the initial threshold. In particular, a sum of the local
surrogate costs in Eq.~\eqref{eq:local} is not in general the cost of
this global candidate. Subsequent searches seek an improvement on it. When
the window advances, the remaining part of the previous trajectory can
be shifted and completed to form a new candidate. A good initial candidate can
improve practical search performance, although it does not change the
worst-case $\mathcal O(\sqrt{N})$ query complexity of Grover-adaptive-search.

From a theoretical perspective, QuantGraph links quantum optimization with practical control. While we have focused on linearized dynamics described by Eq.~\eqref{LTI}, the local search could be extended to nonlinear dynamics by reducing polynomial approximations of the local objective to quadratic form using auxiliary binary variables~\cite{dattani2019quadratizationdiscreteoptimizationquantum}, albeit at higher circuit cost. Another direction is to incorporate quantum algorithms for continuous optimization, which would mitigate the quantization errors inherent in the current discretization scheme \cite{jin2025fixed}. A detailed noise analysis would also clarify robustness to both sensor noise and quantum hardware imperfections. Importantly, feedback and repeated optimization over a short control window highlight how classical control methods can bolster quantum computing primitives, just as quantum-inspired algorithms have become a source of new ideas in classical computing (e.g.~see Ref.~\cite{arrazola2020} and references therein). Advances in control can guide the development of quantum algorithms, and quantum resources may in turn strengthen the performance of established control methods~\cite{chella2022quantum}.

The receding-horizon structure of QuantGraph also offers practical advantages on quantum hardware. Optimizing over a short control horizon $N_c$ reduces the number of decision qubits. Circuit depth and CNOT count also depend on the objective terms, arithmetic precision and, for nonlinear dynamics, the prediction horizon $N_p$ (Appendix~\ref{oracle}). At each sliding-window iteration, the quantum register is reinitialized. The model-predictive-control loop then uses the updated state to respond to deviations caused by earlier control decisions. The threshold $\tau$ decreases whenever an improved candidate is found.

Other quantum optimization methods could also be used within each
control window. Grover-adaptive-search increases the probability of
measuring a candidate that improves on the current cost threshold.
Adiabatic optimization and quantum annealing instead encode the cost
as an energy and seek states of low energy~\cite{farhi2000quantum,kadowaki1998quantum}. For the quadratic
binary objective in Eq.~\eqref{eq:qubo}, replacing each $b_i$ by
$(I-Z_i)/2$, where $Z_i$ is the Pauli $Z$ operator on qubit $i$, gives
a diagonal Ising Hamiltonian with energies equal to the candidate costs.
The time required for adiabatic evolution that follows the ground state depends in part on the
smallest gap between the ground and excited states during the evolution. Practical
implementations must also account for the precision of the Hamiltonian
coefficients and the additional qubits needed to represent its
interactions on hardware with limited connectivity. Constraints
generally require energy penalties or additional variables.

Dissipative methods offer another possibility: controlled interactions
with an environment can drive the system towards states of low
energy~\cite{verstraete2009quantum,ding2024single}. The target state is
made stationary under these dynamics, which must also drive the
initial state towards the desired state or subspace. Convergence
depends on the relaxation properties of the Lindblad generator,
including its spectral gap, and performance also depends on the cost
of implementing the required interactions. For optimization, it may suffice to
find a feasible candidate below a chosen cost threshold. The time needed
to reach this energy range with sufficient probability depends on the
threshold as well as on the evolution or relaxation dynamics.
These methods could be combined with QuantGraph's local
and global stages and its receding-horizon structure. Their performance
would require a separate analysis, since the query bounds in
Section~\ref{sec:analysis} apply specifically to Grover-adaptive-search.

As quantum hardware continues to improve, we anticipate that QuantGraph and similar hybrid algorithms will find increasing applications in real-world optimization problems ranging from autonomous-vehicle trajectory planning to supply-chain optimization and energy-grid management~\cite{gachnang2022quantum, blenninger2024q}. By carefully integrating quantum primitives with classical frameworks and using feedback to update the planned controls, we can develop algorithms that deliver meaningful speedups. This work shows that the intersection of quantum computing, optimization, and control offers promising avenues for developing methods that combine the strengths of each field to address increasingly complex real-world problems~\cite{preskill2018quantum}.

\section{Methods}\label{sec:methods}

\subsection{Problem formulation}

QuantGraph considers the linear time-invariant discrete dynamics of Eq.~\eqref{LTI}, with the states $\mathbf x_k\in\mathbb{R}^{n_x}$ and controls $\mathbf u_k\in\mathbb{R}^{n_u}$. There are different costs driving each stage of the algorithm:
\begin{itemize}
    \item \textbf{Local stage}: The solver optimizes one-step transitions that provide an initial global candidate. Hence the associated cost that provides a surrogate objective for this task is given by,
\begin{equation}
\begin{aligned}
    \quad\quad \ell_{local,k} =& (\mathbf x_k - \mathbf x_k^{\mathrm{ref}})^{\top} \mathbf Q (\mathbf x_k - \mathbf x_k^{\mathrm{ref}}) \\
    &+ (\mathbf u_k - \mathbf u_k^{\mathrm{ref}})^{\top} \mathbf R (\mathbf u_k - \mathbf u_k^{\mathrm{ref}})\\
    &+ (\mathbf x_{k+1} - \mathbf x_{T}^{\mathrm{ref}})^{\top} \mathbf P (\mathbf x_{k+1} - \mathbf x_{T}^{\mathrm{ref}})
\end{aligned}
\label{eq:local}
\end{equation}

\item \textbf{Global stage}: The solver minimizes the cost given by Eq.~\eqref{eq:accumulated_cost} and more specifically the model-predictive-control objective in Eq.~\eqref{eq:MPC_objective}.

\end{itemize}

Each control component is encoded using $M$ bits.
For the scalar controls considered below, this gives a binary vector
$\mathbf b_k\in\{0,1\}^{M}$.
Stacking all control inputs across time produces
$$
    \mathbf b=\bigl[\mathbf b_0^{\!\top},\dots, \mathbf b_{N_c-1}^{\!\top}\bigr]^{\!\top}
      \in\{0,1\}^{MN_c},
$$
and straightforward algebra collects the cumulative cost in the quadratic-unconstrained-binary-optimization form
\begin{equation}
    L(\mathbf b)= \mathbf b^{\top} \mathbf Q \mathbf b + \mathbf q^{\top} \mathbf b + c ,
    \label{eq:qubo}
\end{equation}
with $\mathbf Q\in\mathbb{R}^{MN_c\times MN_c}$ and $\mathbf q\in\mathbb{R}^{MN_c}$. For linear time-invariant systems, these matrices can be efficiently synthesized (in polynomial time) at runtime based on the current state and system dynamics.

\subsection{Two-stage search architecture}

The optimization objective stated in Eq.~\eqref{eq:qubo} proceeds in two stages as seen in Algorithm \ref{alg:quantgraph}.

\paragraph*{\textbf{Local stage}.} The horizon is split into \(T\) single-step sub-problems, each using \(M_{\mathrm{loc}}\) decision qubits. Algorithm \ref{alg:grovermin} is executed on every sub-problem, while the global cost of the resulting candidate sets the initial threshold \(\tau_0\) for the subsequent global search.

\begin{algorithm}[t!]
  \caption{\textsc{QuantGraph}}
  \label{alg:quantgraph}
  \SetAlgoLined
  \SetNlSty{}{}{\hspace{-1em}}   
  \DontPrintSemicolon        

  \KwIn{Initial state $\mathbf x_0$; total horizon $T$, control horizon $1\le N_c\le T$,
        \#bits~$(M_{\mathrm{loc}},M_{\mathrm{glob}})$; search budgets $(\mathcal Q_{\mathrm{loc}},\mathcal Q_{\mathrm{glob}})$}
  \KwOut{Planned state trajectory $\mathbf X^\star$, control sequence $\mathbf U^\star$;}

  \textbf{Setup:}
  Discretize the control function into $2^{M_{\mathrm{loc}}}$ (local) and
  $2^{M_{\mathrm{glob}}}$ (global) binary symbols.

  \textbf{Local Stage}\Comment{Local Warm-Start}\;
  \For{$k\gets0$ \KwTo $T-1$}{
        Form $(\mathbf Q,\mathbf q,c)$ for Eq.~\eqref{eq:local} at $\mathbf x_k$ and choose an admissible initial string $\overline{\mathbf b}_k$\;
        $\begin{aligned}
          &(\mathbf b_k,\ell_k)\gets\\
          &\quad\textsc{GroverMin}\!\bigl(\mathbf Q,\mathbf q,c,M_{\mathrm{loc}},\overline{\mathbf b}_k,\mathcal Q_{\mathrm{loc}}\bigr)
        \end{aligned}$\;
        Decode $\mathbf b_k$ to $\mathbf u_k^{\mathrm{loc}}$\;
        $\mathbf x_{k+1}\gets \mathbf A \mathbf x_k + \mathbf B \mathbf u_k^{\mathrm{loc}}$ \Comment{Linear-time-invariant dynamics only}
  }

  \textbf{Global Stage (Receding-Horizon Grover-adaptive-search).}\;
  $t\gets0$;\; $\mathbf X^\star\gets[\mathbf x_0]$;\; $\mathbf U^\star\gets[\;]$;\;
  $\mathbf x\gets\mathbf x_0$\;
  Round the first $N_c$ local controls to the nearest global grid values and encode them as $\overline{\mathbf b}$\;
  \While{$t\le T-N_c$}{
       Synthesize $(\mathbf Q,\mathbf q,c)$ from Eq.~\eqref{eq:MPC_objective} starting at state $\mathbf{x}$\;
       $\begin{aligned}
         &(\mathbf b^\star,\ell_{\min})\gets\\
         &\quad\textsc{GroverMin}\!\bigl(\mathbf Q,\mathbf q,c,M_{\mathrm{glob}}N_c,\overline{\mathbf b},\mathcal Q_{\mathrm{glob}}\bigr)
       \end{aligned}$\;
        Decode first block of $\mathbf b^\star$ to $\hat{\mathbf{u}}_t$.\;
       $\mathbf{x}\gets \mathbf{f}(\mathbf{x}, \hat{\mathbf{u}}_t)$ \Comment{Apply first input}\;
       Append $\hat{\mathbf{u}}_t$ to $\mathbf{U}^\star$ and $\mathbf{x}$ to $\mathbf{X}^\star$;\;
       Form $\overline{\mathbf b}$ by shifting $\mathbf b^\star$ and repeating its last control block\;
       $t\gets t+1$;\;
}
Apply the remaining $N_c-1$ controls of $\mathbf b^\star$, propagating the dynamics and appending to $\mathbf U^\star$ and $\mathbf X^\star$\;

\textbf{Stage 3 (Return).}\;
\Return{$\mathbf{X}^\star, \mathbf{U}^\star$}\;
\end{algorithm}

\begin{algorithm}[t!]
  \caption{\textsc{GroverMin}$(\mathbf Q,\mathbf q,c,m,\overline{\mathbf b},\mathcal Q)$}
  \label{alg:grovermin}
  \SetAlgoLined
  \SetNlSty{}{}{\hspace{-1em}}
  \SetInd{0.0em}{0.5em}

  \KwIn{Quadratic unconstrained optimization coefficients $(\mathbf Q,\mathbf q,c)$; qubit count $m$; admissible initial string $\overline{\mathbf b}$; integer query budget $\mathcal Q\ge1$}
  \KwOut{Best bitstring found $\mathbf{b}^\star$ and cost $L(\mathbf{b}^\star)$}

  Construct $\widehat L$ using Eq.~\eqref{eq:integer_qubo_objective}\;
  $\mathbf b^\star\gets\overline{\mathbf b}$, $\widehat\tau\gets\widehat L(\mathbf b^\star)$\;
  $K\gets1$, $n_q\gets1$\;

  \While{$n_q<\mathcal Q$}{
 Draw $k\sim\mathcal U\{0,\ldots,\lceil K\rceil-1\}$\;
 \lIf{$n_q+k+1>\mathcal Q$}{\textbf{break}}
 Initialize uniform superposition on $m$ qubits and zero the auxiliary registers\;
      Construct implicit oracle $O_\tau$ that evaluates $\widehat L(\mathbf b)$ and flips phase if $\widehat L(\mathbf b)<\widehat\tau$.\;
 Apply $(D O_\tau)^k$\;
 Measure register to obtain $\mathbf{b}_{\mathrm{cand}}$\;
 $\widehat\ell_{\mathrm{cand}}\gets\widehat L(\mathbf b_{\mathrm{cand}})$ \Comment{Evaluate classically}\;
 $n_q\gets n_q+k+1$\;
 \If{$\widehat\ell_{\mathrm{cand}}<\widehat\tau$}{
 $\widehat\tau\gets\widehat\ell_{\mathrm{cand}}$\;
 $\mathbf{b}^\star\gets \mathbf{b}_{\mathrm{cand}}$\;
 $K\gets1$\;
 }\Else{
 $K\gets\min(6K/5,2^{m/2})$\;
 }
  }
  \Return{$\mathbf b^\star,L(\mathbf b^\star)$}\;
\end{algorithm}

\paragraph*{\textbf{Receding-horizon global stage}.} At control instant \(t\), starting from state $\mathbf{x}_t$, the linear time-invariant dynamics (see Eq.~\eqref{LTI}) and quadratic costs over the prediction horizon $N_p$, with $N_c$ independent controls and the fixed tail, are synthesized into a quadratic-unconstrained-binary-optimization formulation (Eq.~\eqref{eq:qubo}). Grover-adaptive-search is then applied using an implicit oracle derived from this quadratic unconstrained optimization problem definition (Algorithm \ref{alg:grovermin}), searching the space of \(2^{M_{\mathrm{glob}}N_c}\) control sequences with the incumbent bound $\widehat\tau_t$. Only the first element of the current best sequence is applied to the plant, after which the window shifts and the process repeats. The required register therefore never exceeds $M_{\mathrm{glob}}N_c$ plus the ancillary qubits required for the quantum arithmetic in the implicit oracle. Importantly, this procedure can handle trajectories of arbitrary temporal lengths.

\subsection{Grover adaptive search}

As seen in Algorithm \ref{alg:grovermin},
Grover-adaptive-search maintains a classical threshold $\widehat\tau$ and iterates:

\begin{enumerate}[label=(\alph*)]
  \item \textbf{Implicit oracle construction.}
 Constructs a quantum circuit (the implicit oracle) $\mathcal{O}_\tau$ that evaluates the quadratic-unconstrained-binary-optimization's cost function in superposition and performs a conditional phase flip, i.e., $\lvert \mathbf{b}\rangle \mapsto (-1)^{[\widehat L(\mathbf{b})<\widehat\tau]}\lvert \mathbf{b}\rangle$. This requires quantum arithmetic circuits to compute $\widehat L(\mathbf b)$ and a quantum comparator (see Appendix~\ref{oracle} for the explicit circuit construction and logical resource requirements).
  \item \textbf{Randomized amplification.}
        Draws $k\sim\mathcal{U}\{0,\dots,\lceil K\rceil-1\}$ and applies $k$ Grover iterations $G=D\mathcal{O}_\tau$, where \(D=2\lvert s\rangle\langle s\rvert - I\) and $\lvert s\rangle$ is the uniform superposition. Each attempt starts from $\lvert s\rangle$. Failure to improve the cost increases $K$ to $\min(6K/5,2^{m/2})$~\cite{Gilliam_2021,durr1996minimum}.

  \item \textbf{Measurement and update.}
        If the integer candidate cost is below~$\widehat\tau$, updates $\widehat\tau$ and records the index. An improvement resets $K$ to one. The loop stops at the prescribed query budget and retains the best candidate found.
\end{enumerate}

The budget $\mathcal Q$ counts phase oracle calls and classical evaluations of $\widehat L$, including initialization. For exact logical operations, a sufficiently large constant multiple of $2^{m/2}$ gives a fixed probability of finding a minimum of $\widehat L$ per search~\cite{durr1996minimum}. The query bounds below use this convention.

\section{Computational Analysis}\label{sec:analysis}

We analyze the computational advantages of the QuantGraph framework, focusing on its query-complexity scaling and the impact of the quantum-search advantage on the control grid resolution of discretized continuum problems.

\subsection{Query complexity and scalability}

The query complexity of the proposed solver is determined by the combination of the Grover-adaptive-search algorithm and the model-predictive-control architecture~\cite{nayak1999quantum}. We compare this complexity to a monolithic application of Grover-adaptive-search (``original Grover-adaptive-search'') over the entire optimization horizon $T$.

Recall that the control input at each step is discretized using $M$ bits ($M_{\mathrm{glob}}$ in the global stage), and the predictive horizon has length $N_p = T - t$, where the control-input is optimized only for $N_c$ timesteps (the control horizon).

\paragraph*{\textbf{Monolithic search} (original Grover-adaptive-search).}
If a single Grover search were applied to optimize the entire trajectory simultaneously, the total search-space size would be $N_{\mathrm{vanilla}} = (2^M)^T = 2^{MT}$. The query complexity, benefiting from the quadratic speedup, is:
\begin{equation}
    O(\sqrt{N_{\mathrm{vanilla}}}) = O(\sqrt{2^{MT}}) = O(2^{MT/2}).
    \label{eq:complexity_vanilla}
\end{equation}
This scales exponentially with the total horizon $T$, making it intractable for long-duration tasks and demanding significant quantum resources.

\paragraph*{\textbf{QuantGraph framework}.}
QuantGraph operates in a receding-horizon manner. At each time step, the global stage optimizes over the window $N_c$. The search-space size per step is $N_{\mathrm{step}} = 2^{MN_c}$. The query complexity for the Grover-adaptive-search algorithm at each step is $O(\sqrt{N_{\mathrm{step}}}) = O(2^{MN_c/2})$.

The model-predictive-control loop runs for $T-N_c+1$ iterations. The remaining $N_c-1$ controls complete the final plan without further searches. Therefore, the total query complexity for the global stage is:
\begin{equation}
    O((T-N_c+1) \cdot 2^{MN_c/2}) \approx O(T \cdot 2^{MN_c/2}).
    \label{eq:complexity_quantgraph}
\end{equation}
The local stage adds $O(T\,2^{M_{\mathrm{loc}}/2})$ queries.

Crucially, the query complexity varies exponentially with the moving horizon $N_c$. However, for fixed $N_c$ and $M$, the query complexity scales\textit{linearly} with the total time $T$. This shift from exponential to linear scaling in $T$ is the primary computational advantage of the model-predictive-control framework, allowing QuantGraph to handle large $T$ without incurring the exponential cost associated with monolithic search. It is important to note that the cost of implementing these queries also depends on the oracle arithmetic, quantum hardware, error correction and qubit coherence times \cite{montanaro2016quantum}.

\subsection{Precision in discretized continuum problems}

When continuous control problems (e.g., cart pole or double integrator) are discretized into a graph, the precision refers to the control grid resolution. The precision is inherently linked to the discretization granularity $M$~\cite{szablowski2021understanding}. A larger $M$ yields a finer resolution $\Delta u \propto 1/2^M$, reducing the normalized grid spacing $\epsilon$, defined as $\Delta u$ divided by the width of the admissible input interval.

We analyze the achievable precision compared to the classical case (e.g., dynamic programming) for a fixed query budget $\mathcal{Q}$.
The input range, horizon $H$ and success probability are fixed, and $\mathcal Q$ is the budget for one search.

Classical search requires $O(N)$ queries, while quantum search requires $O(\sqrt{N})$. For a budget $\mathcal{Q}$, the maximum searchable-space sizes are:
\begin{equation}
    N_{\mathrm{classical}} \approx \mathcal{Q}, \qquad N_{\mathrm{quantum}} \approx \mathcal{Q}^2.
\end{equation}

We relate the search space size $N$ back to the bits of precision $M$ over a horizon $H$ (where $H$ could be $T$ for monolithic search or $N_c$ for receding-horizon formulation): $N = 2^{MH}$.

\begin{align}
    2^{M_{\mathrm{classical}}H} &\approx \mathcal{Q}, \\
    2^{M_{\mathrm{quantum}}H} &\approx \mathcal{Q}^2.
\end{align}

Taking the logarithm:
\begin{align}
    M_{\mathrm{classical}}H &\approx \log_2(\mathcal{Q}), \\
    M_{\mathrm{quantum}}H &\approx \log_2(\mathcal{Q}^2) = 2 \cdot \log_2(\mathcal{Q}).
\end{align}

Therefore, we find that $M_{\mathrm{quantum}} \approx 2 \cdot M_{\mathrm{classical}}$. For a fixed number of queries, the quantum approach allows for \textbf{twice the number of bits of precision} in the control discretization compared to the classical approach.

This translates into a quadratic reduction in the normalized control grid spacing $\epsilon$,
\begin{equation}
\epsilon_{\mathrm{quantum}} \propto \frac{1}{2^{M_{\mathrm{quantum}}}} \approx \left(\frac{1}{2^{M_{\mathrm{classical}}}}\right)^2 \propto (\epsilon_{\mathrm{classical}})^2.
\label{eq:precision_quadratic}
\end{equation}
This gives a finer control grid at a fixed query budget. Trajectory and cost accuracy also depend on the dynamics, control horizon and arithmetic precision.

\section*{Acknowledgements}
P.V. is supported by the United States Army Research Office under Award No. W911NF-21-S-0009-2. A.P is supported by the J.P. Morgan AI Research Ph.D. Fellowship and the University of Oxford’s Clarendon Fund. M.T.M. is supported by a Royal Society University Research Fellowship. N.A. acknowledges support from the European Research Council (grant agreement 948932) and the Royal Society (URF-R1-191150).

\section*{Author Contributions}
P.V and A.P contributed equally to this work. P.V and A.P performed the simulations and wrote the code under the supervision of I.H, N.A, M.T.M and D.A.S. P.V and A.P wrote the initial draft of manuscript that was iteratively refined under the supervision of I.H, N.A, M.T.M and D.A.S. 

\section*{Declaration of Interests}
The authors declare no competing interests.

\section*{Data Availability}
Illustrative code for the numerical examples is publicly available~\cite{quantgraph_code}.
Requests for further information or data should be sent to the authors.

\appendix

\section{Explicit oracle construction}
\label{oracle}

For the linear dynamics and quadratic costs in Section~\ref{sec:methods},
both stages of QuantGraph take the form of a quadratic unconstrained
binary optimization (QUBO) problem, Eq.~\eqref{eq:qubo}. Here, we describe
the quantum circuit that evaluates this objective and marks control
sequences with a cost below the current threshold.
We use $(\mathbf Q,\mathbf q,c)$ for the coefficients of the corresponding QUBO in either stage.

\subsection{Local and global QUBO problems}
\label{app:stage_qubos}

For the scalar controls used in our numerical demonstrations, we encode
$\mathbf u_k$ at local step $k$ using $M_{\mathrm{loc}}$ bits.
Substituting this encoding and the dynamics
$\mathbf x_{k+1}=\mathbf A\mathbf x_k+\mathbf B\mathbf u_k$ into the
local objective $\ell_{local,k}$ in Eq.~\eqref{eq:local} gives its
binary form

\begin{equation}
\begin{aligned}
    L_k^{\mathrm{loc}}(\mathbf b_k)
    &=
    \mathbf b_k^{\top}\mathbf Q\mathbf b_k
    +(\mathbf q)^{\top}\mathbf b_k
    +c,\\
    &\mathbf b_k\in\{0,1\}^{M_{\mathrm{loc}}}.
\end{aligned}
    \label{eq:local_stage_qubo}
\end{equation}

The state $\mathbf x_k$ remains fixed during this search.

At MPC step $t$, with $1\le N_c\le T-t$, we collect the bits for the
$N_c$ optimized controls into
\begin{equation}
    \mathbf b_t^{\mathrm{glob}}
    =
    \begin{bmatrix}
        \mathbf b_t^{\top} &
        \cdots &
        \mathbf b_{t+N_c-1}^{\top}
    \end{bmatrix}^{\top}
    \in\{0,1\}^{M_{\mathrm{glob}}N_c}.
    \label{eq:global_decision_string}
\end{equation}
Substituting these controls into Eq.~\eqref{eq:MPC_objective}, with the
last input held fixed beyond the control horizon, gives the global
objective over the remaining predictive horizon $N_p=T-t$,
\begin{equation}
    L_t^{\mathrm{glob}}(\mathbf b_t^{\mathrm{glob}})
    =
    (\mathbf b_t^{\mathrm{glob}})^{\top}
        \mathbf Q\mathbf b_t^{\mathrm{glob}}
    +(\mathbf q)^{\top}
        \mathbf b_t^{\mathrm{glob}}
    +c.
    \label{eq:global_stage_qubo}
\end{equation}

The coefficients in both stages are determined by the current state,
dynamics, reference trajectory, and cost matrices, as described in
Section~\ref{sec:methods}.

The same arithmetic applies to either stage. Omitting the stage index,
we write
\begin{equation}
    L(\mathbf b)
    =
    \mathbf b^{\top}\mathbf Q\mathbf b
    +\mathbf q^{\top}\mathbf b+c
    =
    c+\sum_{i=1}^{m}\alpha_i b_i
      +\sum_{1\leq i<j\leq m}\beta_{ij}b_ib_j,
    \label{eq:canonical_stage_qubo}
\end{equation}
where $m$ is the number of decision qubits passed to
\textsc{GroverMin} in Algorithm~\ref{alg:grovermin}. As in
Eq.~\eqref{eq:qubo}, $\mathbf Q$ denotes the binary objective matrix
here. It is distinct from the state weighting matrix in
Eq.~\eqref{eq:cost}. The coefficients are
\begin{equation}
    \alpha_i=q_i+Q_{ii},
    \qquad
    \beta_{ij}=Q_{ij}+Q_{ji}.
    \label{eq:canonical_stage_coefficients}
\end{equation}
Here we used $b_i^2=b_i$ to combine the diagonal quadratic terms with
the linear terms.

\subsection{Reversible cost evaluation}
\label{app:reversible_cost_evaluation}
To evaluate the cost with finite precision, we first express it in a
fixed cost unit. We then scale the numerical QUBO coefficients by
$2^p$ and round them to integers, where the nonnegative integer $p$
sets the number of fractional bits,
\begin{equation}
\begin{aligned}
    &\widehat\alpha_i
    =\operatorname{round}(2^p\alpha_i), \quad
    \widehat\beta_{ij}
    =\operatorname{round}(2^p\beta_{ij}),
     \\
    &\widehat c=\operatorname{round}(2^pc).
    \label{eq:oracle_coefficient_quantization}
\end{aligned}
\end{equation}
The resulting integer objective is
\begin{equation}
    \widehat L(\mathbf b)
    =
    \widehat c+\sum_i\widehat\alpha_i b_i
    +\sum_{i<j}\widehat\beta_{ij}b_ib_j.
    \label{eq:integer_qubo_objective}
\end{equation}

The threshold must use the same cost representation as the candidates.
For the best candidate found so far, $\overline{\mathbf b}$, we therefore set
\begin{equation}
    \widehat\tau=\widehat L(\overline{\mathbf b}).
    \label{eq:consistent_integer_threshold}
\end{equation}
For a separately specified threshold, we use
$\widehat\tau=\operatorname{round}(2^p\tau)$. Rounding can
change the ordering of nearby costs, so an improvement in
$\widehat L$ need not improve the unrounded objective. This precision
is separate from the $M_{\mathrm{loc}}$ or $M_{\mathrm{glob}}$ bits used
to represent each control.

Let register $\mathsf B$ hold the $m$ decision bits and let
$\mathsf V$ hold the signed cost
difference, using $m_{\mathrm{val}}$ qubits in two's complement
representation. Register $\mathsf W$ holds the temporary values needed
for the arithmetic. The unitary that adds the cost difference to
$\mathsf V$ is
\begin{equation}
\begin{aligned}
    &U_{\tau}
    \ket{\mathbf b}_{\mathsf B}\ket{z}_{\mathsf V}\ket{0}_{\mathsf W} \\
    &\hspace{2cm}=
    \ket{\mathbf b}_{\mathsf B}
    \ket{
        \bigl(z+\widehat L(\mathbf b)-\widehat\tau\bigr)
        \bmod 2^{m_{\mathrm{val}}}
    }_{\mathsf V}
    \ket{0}_{\mathsf W}.
    \label{eq:reversible_value_unitary}
\end{aligned}
\end{equation}

To construct $U_{\tau}$, we add $\widehat c-\widehat\tau$ to $\mathsf V$,
then add each $\widehat\alpha_i$ when $b_i=1$ and each
$\widehat\beta_{ij}$ when both $b_i=b_j=1$.

Negative coefficients use the same two's complement representation.
The additions can be implemented with reversible integer adders.
Alternatively, we apply a quantum Fourier transform to the value
register, perform the required phase rotations, and apply the inverse
transform~\cite{Gilliam_2021}. When this register starts in zero,
Hadamard gates prepare the required initial Fourier state.

The value register must be wide enough to avoid overflow, which could
give the cost difference the wrong sign. For a given QUBO and threshold,
define
\begin{equation}
    R_{\widehat\tau}
    =
    \left|\widehat c-\widehat\tau\right|
    +\sum_i\left|\widehat\alpha_i\right|
    +\sum_{i<j}\left|\widehat\beta_{ij}\right|.
    \label{eq:oracle_value_bound}
\end{equation}
The triangle inequality gives
\begin{equation}
    \left|
      \widehat L(\mathbf b)-\widehat\tau
    \right|
    \leq R_{\widehat\tau},
\end{equation}
for every decision string. A sufficient register width is therefore
\begin{equation}
    m_{\mathrm{val}}
    \geq
    1+\left\lceil
        \log_2\!\left(R_{\widehat\tau}+1\right)
      \right\rceil.
    \label{eq:oracle_value_register_width}
\end{equation}
A tighter interval for the cost difference can reduce this width.
Our implementation combines coefficient bounds in the binary and Ising
representations, then uses the smallest signed register that contains
their intersection.

\subsection{Local and global phase oracles}
\label{app:local_global_phase_oracles}
The oracle first computes the cost difference, flips the phase when
that difference is negative, and reverses the computation to reset the
auxiliary registers.
Let $Z_{\mathrm{sgn}}$ denote a Pauli $Z$ acting on the most significant,
or sign, qubit of the value register. For the local QUBO in
Eq.~\eqref{eq:local_stage_qubo}, let
$U_{k,\tau}^{\mathrm{loc}}$ denote the unitary in
Eq.~\eqref{eq:reversible_value_unitary} constructed from
$(\mathbf Q,\mathbf q,
c)$. The local phase oracle is
\begin{equation}
    \mathcal O_{k,\tau}^{\mathrm{loc}}
    =
    \left(U_{k,\tau}^{\mathrm{loc}}\right)^{\dagger}
    \left(
        I_{\mathsf B}\otimes Z_{\mathrm{sgn}}\otimes I_{\mathsf W}
    \right)
    U_{k,\tau}^{\mathrm{loc}}.
    \label{eq:explicit_local_oracle}
\end{equation}
When the register width satisfies
Eq.~\eqref{eq:oracle_value_register_width}, it acts as
\begin{equation}
    \mathcal O_{k,\tau}^{\mathrm{loc}}
    \ket{\mathbf b_k}_{\mathsf B}\ket{0}_{\mathsf V}\ket{0}_{\mathsf W}
    =
    (-1)^{
      [\widehat L_k^{\mathrm{loc}}(\mathbf b_k)<\widehat\tau]
    }
    \ket{\mathbf b_k}_{\mathsf B}\ket{0}_{\mathsf V}\ket{0}_{\mathsf W}.
    \label{eq:explicit_local_oracle_action}
\end{equation}

Similarly, let $U_{t,\tau}^{\mathrm{glob}}$ be constructed from
$(\mathbf Q,\mathbf q,
c)$ in Eq.~\eqref{eq:global_stage_qubo}. The global phase oracle is
\begin{equation}
    \mathcal O_{t,\tau}^{\mathrm{glob}}
    =
    \left(U_{t,\tau}^{\mathrm{glob}}\right)^{\dagger}
    \left(
        I_{\mathsf B}\otimes Z_{\mathrm{sgn}}\otimes I_{\mathsf W}
    \right)
    U_{t,\tau}^{\mathrm{glob}}.
    \label{eq:explicit_global_oracle}
\end{equation}
Its action is
\begin{equation}
\begin{aligned}
    \mathcal O_{t,\tau}^{\mathrm{glob}}
    \ket{\mathbf b_t^{\mathrm{glob}}}_{\mathsf B}
    \ket{0}_{\mathsf V}\ket{0}_{\mathsf W}
    &=
    (-1)^{
      [\widehat L_t^{\mathrm{glob}}
        (\mathbf b_t^{\mathrm{glob}})<\widehat\tau]
    }
    \\
    &\quad\times
    \ket{\mathbf b_t^{\mathrm{glob}}}_{\mathsf B}
    \ket{0}_{\mathsf V}\ket{0}_{\mathsf W} .
\end{aligned}
\label{eq:explicit_global_oracle_action}
\end{equation}
Both oracles therefore mark control strings whose quantized cost is
strictly lower than the threshold, following the acceptance rule in
Algorithm~\ref{alg:grovermin}.

The constant $c$ does not affect which candidate minimizes the cost.
Algorithm~\ref{alg:grovermin} includes $\widehat c$ in both costs and threshold.
Alternatively, we may remove it and shift the threshold by
$\widehat\tau\leftarrow\widehat\tau-\widehat c$.

\subsection{Grover iteration and logical resources}
\label{app:oracle_resources}
We can also perform the search while storing each cost difference
alongside its control string. Applying Hadamard gates to the decision
register and then evaluating the cost gives the preparation operator
\[
    A_{\tau}
    =U_{\tau}
    \left(H^{\otimes m}\otimes I_{\mathsf V}\otimes I_{\mathsf W}\right).
\]
Its action on the initial registers is
\begin{equation}
\begin{aligned}
    &A_{\tau}\ket{0}_{\mathsf B}\ket{0}_{\mathsf V}\ket{0}_{\mathsf W}\\
    &\quad=
    2^{-m/2}
    \sum_{\mathbf b\in\{0,1\}^{m}}
    \ket{\mathbf b}_{\mathsf B}
    \ket{\widehat L(\mathbf b)-\widehat\tau}_{\mathsf V}
    \ket{0}_{\mathsf W}.
\end{aligned}
    \label{eq:oracle_state_preparation}
\end{equation}
The signed value in the ket uses the modular encoding in
Eq.~\eqref{eq:reversible_value_unitary}.
On the full register, the marking operation is
$S_{\mathrm{sgn}}=I_{\mathsf B}\otimes
Z_{\mathrm{sgn}}\otimes I_{\mathsf W}$.
The reflection is taken about the prepared state, giving the
Grover iteration
\begin{equation}
    G_{\tau}
    =
    A_{\tau}S_0A_{\tau}^{\dagger}S_{\mathrm{sgn}},
    \qquad
    S_0=I-2\ket{0}\!\bra{0},
    \label{eq:oracle_grover_iterate}
\end{equation}
where $\ket{0}$ in $S_0$ denotes the joint zero state of
$\mathsf B$, $\mathsf V$, and $\mathsf W$.
For initially empty auxiliary registers, applying $U_{\tau}$ before
$G_{\tau}$ and $U_{\tau}^{\dagger}$ afterwards gives the decision
register form $G=D\mathcal O_{\tau}$, up to a global phase.

Counting the decision, value, and
work registers, the two stages require
\begin{align}
    n_{\mathrm{logical}}^{\mathrm{loc}}
    &=
    M_{\mathrm{loc}}
    +m_{\mathrm{val}}^{\mathrm{loc}}
    +a_{\mathrm{work}}^{\mathrm{loc}},
    \\
    n_{\mathrm{logical}}^{\mathrm{glob}}
    &=
    M_{\mathrm{glob}}N_c
    +m_{\mathrm{val}}^{\mathrm{glob}}
    +a_{\mathrm{work}}^{\mathrm{glob}},
    \label{eq:local_global_logical_qubits}
\end{align}
logical qubits for scalar controls. The workspace size $a_{\mathrm{work}}$
counts the qubits in $\mathsf W$ and depends on the arithmetic
implementation. For $n_u$ control components, the decision register
contains $M_{\mathrm{loc}}n_u$ or $M_{\mathrm{glob}}N_cn_u$ qubits.
If $s_1$ and $s_2$ count the
nonzero linear and quadratic terms in Eq.~\eqref{eq:canonical_stage_qubo},
one cost evaluation in the Fourier basis requires
\begin{equation}
    O\!\left(
        m_{\mathrm{val}}(1+s_1+s_2)
        +m_{\mathrm{val}}^2
      \right)
    \label{eq:oracle_gate_complexity}
\end{equation}
phase rotations, including those in the forward and inverse quantum
Fourier transforms. This count
treats the rotations as exact logical gates. Their decomposition into
a fixed gate set adds a cost that depends on the required accuracy.
A phase oracle call evaluates the cost once and then reverses that
evaluation to reset the auxiliary registers. Preparing the decision
superposition and implementing the Grover reflection incur separate
gate costs.

\subsection{Nonlinear trajectory costs}
\label{app:nonlinear_oracle}
For nonlinear dynamics, the objective need not be quadratic in the
control bits. We can instead evaluate the trajectory directly using
an approximation $\widehat f$ to the dynamics $f$ in
Eq.~\eqref{nonlinear_dynamics} at the chosen arithmetic precision.
Starting from $\mathbf x_t$, we compute each predicted state into a
separate auxiliary register and retain the preceding state and control.
This allows us to reverse the calculation even when the state update
is not invertible. We then evaluate the stage costs $\ell_k$ and
terminal cost $\phi$ in Eq.~\eqref{eq:cost}, add their contributions
to the value register, and reverse the intermediate calculations.
Subtracting the threshold gives the operation in
Eq.~\eqref{eq:reversible_value_unitary}, where $\widehat L$ now denotes
the integer trajectory cost. The same phase oracle then marks
improving controls.

This construction requires reversible circuits for all operations
in $\widehat f$ and the cost calculation. Finite precision can change
the ordering of costs, and the value register must avoid overflow. The workspace
and gate count depend on the state dimension, predictive horizon
$N_p$, and arithmetic used to evaluate the nonlinear functions.
These operations are not included in the QUBO estimate in
Eq.~\eqref{eq:oracle_gate_complexity}. We do not give a circuit or
resource estimate for the cart pole here.

\section{Constraint handling}
\label{app:constraints}

In classical dynamic programming, the Bellman recursion minimizes over
the inputs that are admissible at each state~\cite{DP_2010}. Constraints
can therefore be incorporated into the allowed choices without adding
penalties to the objective. In our examples, the system dynamics and
control bounds are satisfied through the way we decode the control
strings and evaluate their trajectories. Additional state and path
constraints require further oracle calculations and are not tested in
these benchmarks.

For each control sequence, the states are obtained by propagating
$\mathbf x_{k+1}=f(\mathbf x_k,\mathbf u_k)$ from the current measured
state $\mathbf x_t$. This formulation is known as single shooting.
The decision string in Eq.~\eqref{eq:global_decision_string} contains
only control bits. Collocation or multiple shooting would
instead introduce state variables and equality constraints for the
dynamics, requiring additional qubits and constraint checks or penalties.

For each control component $i=1,\ldots,n_u$, let $u_{\max,i}$ denote
its allowed magnitude. We satisfy
$|u_{k,i}|\le u_{\max,i}$ by choosing the binary encoding to lie within
this interval. Every decision string then gives an admissible input,
so the oracle needs no separate test for these bounds.

Additional constraints
$g_j(\mathbf x_k,\mathbf u_k)\le 0$, with $j=1,\ldots,n_g$, can also
be evaluated from the decision string. We take each $g_j$ to be
dimensionless, dividing physical constraint residuals by fixed positive
reference scales where needed. This normalization leaves the feasible
set unchanged. For the global search at time
$t$, one approach is to add a penalty for each violation,
\begin{equation}
    L^{\mathrm{pen}}(\mathbf b)
    =
    L(\mathbf b)
    +\lambda\sum_{k=t}^{T-1}\sum_{j=1}^{n_g}
      \max\!\left(0,\,g_j(\mathbf x_k(\mathbf b),\mathbf u_k(\mathbf b))\right)^{2},
    \label{eq:soft_constraint_objective}
\end{equation}
where $\lambda>0$ has the same cost units as $L$. A constraint on the terminal
state can be included through an additional penalty evaluated at
$\mathbf x_T$.
The oracle evaluates each constraint and adds its penalty to the
value register before phase marking and uncomputation, as in
Eqs.~\eqref{eq:explicit_local_oracle} and
\eqref{eq:explicit_global_oracle}. This requires additional reversible
arithmetic and may require workspace beyond the QUBO evaluator in
Appendix~\ref{app:reversible_cost_evaluation}.

The penalty weight must ensure that a violation cannot be compensated
by a lower value of the original cost. For the quantized QUBO objective,
the range is bounded by
\begin{equation}
    \widehat\Delta=\max_{\mathbf b}\widehat L(\mathbf b)-\min_{\mathbf b}\widehat L(\mathbf b) \leq 2R_{\widehat\tau},
\end{equation}
with $R_{\widehat\tau}$ from Eq.~\eqref{eq:oracle_value_bound}.
Let $\widehat\Pi(\mathbf b)$ represent the total penalty as a
nonnegative integer and let $\widehat\lambda$ be its positive integer
weight. The objective used by the oracle is then
$\widehat L^{\mathrm{pen}}(\mathbf b)
=\widehat L(\mathbf b)+\widehat\lambda\,\widehat\Pi(\mathbf b)$.
Suppose a feasible string exists, with $\widehat\Pi=0$ for every
feasible string and $\widehat\Pi\ge1$ for every infeasible string.
Under these assumptions,
\begin{equation}
    \widehat\lambda>\widehat\Delta
    \label{eq:exact_penalty_condition}
\end{equation}
ensures that minimizing $\widehat L^{\mathrm{pen}}$ gives a feasible
optimum on the encoded grid. The penalty representation must retain
enough precision to distinguish a violation from zero. If the current
best candidate is feasible, every string below its penalized cost is
also feasible. The bound does not, however, provide that initial
feasible candidate.

The penalties increase the range of cost differences stored in the
value register. We must therefore update the bound used to choose its
width, as in Eq.~\eqref{eq:oracle_value_register_width}.
The arithmetic also depends on the constraint functions and the
number of predicted states where they are evaluated.
Only $N_c$ controls are independent decisions, but the constraints
may need to be checked over all $N_p=T-t$ predicted steps.

Constraints also affect the number of candidates that the oracle marks.
When the oracle marks only feasible improvements, let $a>0$ be their
fraction among all decision strings. This is the case for an explicit
feasibility test, or for the penalty construction above with a feasible
current candidate. Finding one of these strings
takes $\mathcal O(a^{-1/2})$ Grover iterations on average.
This fraction can be smaller than the fraction of feasible strings
alone. Randomized amplification accommodates an unknown number of
marked strings, but the dependence on $a$ remains. If no improving
candidate exists, the search must stop under a specified rule and
retain the current candidate.

The bound in Eq.~\eqref{eq:exact_penalty_condition} can be conservative.
For a fixed register width, a large penalty may leave fewer bits for
resolving differences between feasible costs. A tighter sufficient
bound can reduce this loss of precision.

\begin{table}[!t]
  \small
  \caption{Numerical settings for the Qiskit implementations. Shot counts are per search attempt. State coordinates
  are ordered as $(q,\dot q)$ for the double integrator and
  $(q,\theta,\dot q,\dot\theta)$ for the cart pole.}
  \label{tab:illustrative_setup}
  \centering
  \setlength{\tabcolsep}{3pt}
  \renewcommand{\arraystretch}{1.1}
  \begin{tabular}{@{}p{0.39\linewidth}p{0.32\linewidth}p{0.23\linewidth}@{}}
    \toprule
    Setting & \mbox{Double integrator} & Cart pole \\
    \midrule
    Time interval [$\mathrm{s}$] & $0.1$ & $0.1$ \\
    Duration [$\mathrm{s}$] & $1.1$ & $6.1$ \\
    Initial state & $(0,0)$ & $(0,\pi,0,0)$ \\
    Target state & $(1,0)$ & $(0,2\pi,0,0)$ \\
    Local control bits & $4$ & Disabled \\
    Global control bits & $3$ & $3$ \\
    Control horizon $N_c$ & $2$ & $6$ \\
    Force interval [$\mathrm{N}$] & $[-5,5]$ & $[-10,10]$ \\
    Local query budget & $64$ & Disabled \\
    Global query budget & $129$ & $3072$ \\
    Local shots & $4$ & Disabled \\
    Global shots & $16$ & $1$ \\
    Input weight $\mathbf R$ & $40$ & $5$ \\
    \bottomrule
  \end{tabular}
\end{table}

We can also test feasibility separately from the cost. Let
$\chi_{\mathrm{feas}}(\mathbf b)$ be a Boolean flag that equals one
exactly when the string satisfies the encoded constraints. The oracle
then marks a candidate when
$[\widehat L(\mathbf b)<\widehat\tau]\wedge
\chi_{\mathrm{feas}}(\mathbf b)$ holds.
This leaves the objective coefficients unchanged, but requires
workspace to evaluate feasibility and a phase operation conditioned on
both tests. Preparing a superposition of feasible strings is another
possibility. Its benefit depends on the cost of preparing that state
and reflecting about it, particularly when feasibility requires
propagating the dynamics.

The arithmetic construction also extends to polynomial objectives of
degree greater than two. A monomial of degree $d$ contributes an
addition conditioned on $d$ decision bits. Such terms can be evaluated
directly, without introducing the auxiliary binary variables needed to
reduce them to quadratic form for an Ising annealer. Extending the
resource estimate in Eq.~\eqref{eq:oracle_gate_complexity} requires
counting both the nonzero monomials and the gates used to implement
their controlled additions.

Finally, an admissible plan in one window does not guarantee that a
feasible plan will exist at every subsequent step. This property,
known as recursive feasibility, requires additional MPC conditions,
often involving a terminal constraint set and a suitable terminal cost.
We have not imposed these conditions here. For a quantum solver,
the controller must also retain a feasible plan when the search finds
no improvement. Establishing recursive feasibility would require
specifying this fallback and accounting for search failures.

\section{Numerical setup}
\label{app:numerical_setup}

Table~\ref{tab:illustrative_setup} summarizes the Qiskit implementations provided in the repository~\cite{quantgraph_code}. The double integrator uses local and global search, whereas the cart pole uses only global search, initialized from a zero control sequence. The upright configuration is $\theta=0$ modulo $2\pi$. Beyond the control horizon, the last optimized input is held fixed as in Eq.~\eqref{eq:MPC_objective}. The control grid contains $2^M$ equally spaced values from $-u_{\max}$ to $u_{\max}-\Delta u$, where $\Delta u=2u_{\max}/2^M$, ensuring that zero force is represented exactly.

The double integrator is propagated exactly under constant input, while cart pole predictions use adaptive Runge--Kutta integration. Each global optimization starts from the current state and predicts over the full remaining task, with inputs held constant between control updates.

The state references are the targets in Table~\ref{tab:illustrative_setup}, with zero reference input. Using Eq.~\eqref{eq:cost}, the state weights are $\mathbf Q=\operatorname{diag}(5\times10^5,5\times10^3)$ and $\operatorname{diag}(30,10,10,1)$ for the double integrator and cart pole, respectively. The corresponding terminal weights are $\mathbf P=\operatorname{diag}(10^6,2\times10^4)$ and $\operatorname{diag}(12500,10,10000,1000)$. The double integrator uses the same weights in both stages. The cart pole running cost further includes $100\sin^2\theta+100(1-\cos\theta)^2$, with the same terms weighted by $10^4$ in the terminal cost. The implementations use Python 3.11 with NumPy, SciPy, Numba and Matplotlib, while Qiskit 2.1.2 and DOCPlex are used for constructing the quadratic objective, simulating the reversible cost evaluation, phase marking and uncomputation using the matrix product state method.

\bibliography{references}

\end{document}